\documentclass[fleqn,usenatbib]{mnras}

\usepackage{newtxtext,newtxmath}

\usepackage[T1]{fontenc}
\usepackage{ae,aecompl}

\usepackage{bigints}
\usepackage{graphicx}	
\usepackage{amssymb}	
\title[]{Testing the rotation versus merger scenario in the galaxy cluster Abell 2107
}
\author[A. Liu et al.]{
Ang Liu,$^{1,2,3}$\thanks{E-mail: liuang@arcetri.astro.it (A. Liu)}
Paolo Tozzi,$^{1}$\thanks{E-mail: ptozzi@arcetri.astro.it (P. Tozzi)}
\\
$^{1}$INAF Osservatorio Astrofisico di Arcetri, Largo E. Fermi, I-50122 Firenze, Italy\\
$^{2}$Department of Physics, Sapienza University of Rome, I-00185 Rome, Italy\\
$^{3}$Department of Physics, University of Rome Tor Vergata, I-00133, Rome, Italy\\
}

\date{Accepted XXX. Received YYY; in original form ZZZ}

\pubyear{2018}

\begin{document}


\maketitle

\begin{abstract}
We search for global rotation of the intracluster medium (ICM) in the galaxy cluster Abell 2107, where
previous studies have detected rotational motion in the member galaxies with a high significance level.
By fitting the centroid of the iron $K_{\alpha}$ line complex at 6.7--6.9 keV rest frame in {\sl Chandra}
ACIS-I spectra, we identify the possible rotation axis with the line that maximizes
the difference between the emission-weighted spectroscopic redshift measured in the two halves
defined by the line itself.  Then, we measure the emission-weighted redshift in linear regions
parallel to the preferred rotation axis, and find a significant gradient as a function of the projected
distance from the rotation axis, compatible with a rotation pattern with maximum tangential velocity
${\tt v}_{\rm max}=1380\pm 600$ km/s at a radius $\lambda_0\sim 160$ kpc.
This result, if interpreted in the framework of hydrostatic equilibrium, as suggested by the regular morphology
of Abell 2107, would imply a large mass correction of the order of $\Delta M = (6 \pm 4)\times 10^{13} M_\odot$
at $\sim 160$ kpc, which is incompatible with the cluster morphology itself.  A more conservative
interpretation may be provided by an unnoticed off-center, head-on collision between two comparable halos.
Our analysis confirms the peculiar dynamical nature of the otherwise regular cluster Abell 2107, but is not
able to resolve the rotation vs merger scenario, a science case that can be addressed by 
the next-generation X-ray facilities carrying X-ray bolometers onboard.
\end{abstract}



\section{Introduction}
Galaxy clusters are the largest virialized structures in the Universe, and have been largely used to map
the growth of cosmic structures through cosmic ages, and, ultimately, to constrain cosmological models
\citep{borgani2008,allen2011} and modified gravity models \citep{schmidt2009,mitchell2018}.
The key quantity is the virialized mass, which is often computed under the assumptions of spherical
symmetry, isotropic velocity distribution of the member galaxies, and hydrostatic equilibrium of the
X-ray emitting intracluster medium (ICM). However, despite virialization is considered a fair assumption,
substantiated by numerical simulations \citep[see][]{2012Kravtsov}, it is now commonly accepted that it is
not completely satisfied in several cases, as suggested by in-depth analysis of hydrodynamical numerical
studies \citep[see, e.g.][]{2016Biffi} and by the increasing claims of dynamical substructures and
bulk motions observed in massive clusters \citep{girardi1997,parekh2015,liu2016}.
In addition to the presence of bulk motions involving significant amount of the cluster mass,
another aspect that can significantly affect mass estimates is the presence of global rotation,
a possibility which is often ignored when applying the classic mass-weighting methods
to galaxy clusters. In dynamical studies the possible presence of global rotation is almost always
neglected, with the exception of few numerical works \citep{2009Lau,2011Biffi}.
From the observational point of view, there are several ongoing projects aiming at quantifying the impact of
bulk motions, particularly associated with mergers in the external regions of clusters
\citep[see, e.g.][and references therein]{2018Ghirardini}, and, in addition, detailed works have been
devoted to the non thermal pressure contribution from the combination of turbulence and bulk motions,
in which rotation can be considered a particular case, despite not explicitly treated
\citep[see][]{eckert2019}.

A fundamental issue concerns the observational difficulty in disentangling asymmetric bulk motions,
due to the presence of substructures with anisotropic velocity distribution associated with multiple
off-centered mergers or continuous accretion of matter along filaments, from a global rotation.
In other words, it is almost impossible to distinguish two overlapping clusters from one rotating
cluster \citep[see][]{oegerle1992}, if not by tracing the rotation curve with high spatial resolution.
In the optical band, these studies are practically limited by the sparse and discrete sampling of the
velocity along the line-of-sight, due to the limited number of member galaxies.  To date,
the number of redshifts available in single clusters can reach a maximum of a thousand
cluster members only for a few, well known targets \citep[see][]{2014Rosati,rines2016}.
Due to these difficulties, dynamical studies in the optical band, based on spectroscopic redshift
distribution of the member galaxies, are mostly focused on the infall of smaller halos in the
outskirts, or galaxies flowing along filamentary structures that are observed to connect the cluster
to the mildly non-linear large scale structures of the Universe.

Although very few systematic studies of global rotation have been done, it was possible to show that
a minority of clusters do show global rotation not associated with recent mergers, on the basis of
spectroscopic samples from SDSS and 2dFGRS \citep{2007Hwang}. Other studies on cluster rotation have
been focused to several specific targets.  Apart from a few suggestions of cluster rotation from
velocity gradients \citep[for a complete set of historical references see][]{kalinkov2005},
SC0316-44 was the first cluster for which a claim of rotation was made
on the basis of the analysis of only 15 member galaxies \citep{1983Materne}. Abell 2107 (hereafter A2107)
has been claimed to show spatial correlations in the galaxy velocities, consistent with rotation,
in a study by \citet{oegerle1992}, based on the redshifts of 68 member galaxies.  However, the
pure rotation model does not account for the peculiar velocity of its brightest cluster galaxy
(BCG) of $\sim 270 $ km/s with respect to the bulk of the cluster.
Eventually, a more in-depth analysis confirmed the rotation of A2107 \citep{kalinkov2005},
with the strongest evidence for rotation being the consistent positional angle of the velocity gradient
for consecutive galaxy subsamples.  Recently, a few systematic studies, using
SDSS DR 9 and DR10 spectroscopic data and new algorithms, found
rotation in a significant but vastly different fraction of their samples, depending on the
sample selection \citep[see][]{2015aTovmassian,2015bTovmassian,2017Manolopoulou}.  If confirmed, a
robust assessment of a widespread global rotation and the associated amount of rotational support
across the cluster population, would allow one to include this contribution to the well known
effects of turbulence and disordered bulk motions, and therefore to alleviate the bias affecting
masses measured through classic hydrostatic equilibrium.  Pushing further ahead with this study,
we also remark that a statistical treatment of rotation among clusters of galaxies can be used to
constrain the origin and growth of angular momentum in cosmic structures at cluster scales.

Another observational window potentially relevant to measure the inner dynamical structure of
clusters, including rotation, is provided by the X-ray emission from the ICM.   This technique
has the advantage of tracing the continuous distribution of collisional matter constituted
by the dominant baryonic component in clusters.  Bulk motions in the ICM can be
detected by measuring the spatial distribution of its emission-weighted redshift, and hence
radial velocity, in limited regions.  This is done by fitting the position of the prominent
iron $K_{\alpha}$ line at 6.7--6.9~keV in the ICM X-ray spectrum. These measurements, however,
require high spectral and spatial resolutions at the same time, due to the expected patchy distribution
of bulk motions.  The tight requirement on spatial resolution limits these studies to the use of CCD data,
which, in turn, implies a poor spectral resolution. Given this limitation, the measurement of the
centroid of the most prominent emission line complex in CCD spectra provides the most accurate
spatially-resolved measurement for the ICM redshift, and the associated statistical error.
The uncertainty on the redshift $\sigma_z$ strongly depends on the strength of the signal, on the
modelization of the thermal structure of the ICM, and on calibration issues, and typically is
found in the range $0.002< \sigma_z< 0.01$ for bright, nearby clusters and groups observed with
{\sl Chandra}. We refer the reader to \citet{2011Yu} for a discussion on the accuracy on the global
ICM redshift and to \citet{liu2015,liu2016} for the accuracy in spatially resolved analysis of the ICM.
Therefore, with a typical uncertainty on ICM velocity in nearby clusters in the range
$c\times \sigma_z/(1+z) \sim 500 - 2000$ km/s, the search of bulk motions provided positive
results only in a few clusters with highly disturbed dynamical status
\citep[see][and references therein]{parekh2015,liu2016}, including the remarkable case of
the Bullet Cluster, where supersonic bulk motion has been identified along the line of sight
and perpendicularly to the merger axis \citep{liu2015}. In general, a direct comparison of bulk
motions along the line of sight in the galaxy members and in the ICM, shows that it is hard to
associate galaxies and ICM motions \citep[see, e.g.][]{liu2018b}.  This is due to the
different dynamical evolution of merger and infall in
collisionless (dark matter and galaxies) and collisional (ICM) components, which appear to become
spatially decoupled within one dynamical time, as we can directly witness in the plane of
the sky in the case of the so called bullet-like clusters. In any case, global rotation, if any, is expected
to be characterized by velocities significantly lower than the typical velocity dispersion in clusters,
which makes it very hard to blindly search for global rotation of the ICM using the available
spectral resolution of CCD X-ray imagers that can at best identify velocity differences of
$\sim 1000$ km/s or larger, as previously mentioned.  As a matter of fact, the study of ICM
bulk motions will be always limited to a few cases until the advent of X-ray bolometers, which
will be on board of XRISM and Athena \citep{2018Guainazzi}. In the case of XRISM, the
low angular resolution of the order of $\sim 1^\prime$ will enable study of bulk motions only
in nearby clusters, since the signal from the ICM in distant clusters will be smeared out.  On the
other hand, the $5-10$ arcsec angular resolution of the bolometer onboard Athena \citep{2018Barret} is expected
to perform to the point of revolutionizing the study of ICM bulk motions.
Currently, the launch of XRISM is planned for the year 2021
\footnote{See https://heasarc.gsfc.nasa.gov/docs/xrism/}, while Athena is expected
to be launched around the year 2031 \footnote{See https://www.the-athena-x-ray-observatory.eu/}.
This impasse in the study of ICM dynamics follows the dramatic loss of the Hitomi satellite
\citep{2018Hitomi}, which was anyhow able to provide a high-spectral resolution view of the Perseus
cluster, showing the effects of turbulence and bulk motions in its core \citep{2016Hitomi}.

In this context, no firm detection of global rotation in the ICM has been reported so far, but a fair number
of numerical studies have been devoted to the investigation of the effects of major, off-centered
mergers and continuous/minor merger infall from filaments on ICM rotation \citep{bianconi2013,baldi2017}.
The only chance to test observationally ICM rotation is to focus on some extreme case, where the
effects of rotation are maximum.  In this work, we make a first attempt to find a signature of rotation
in the ICM, by targeting A2107, that is, to our knowledge, the best candidate for such a study.
A2107 is a massive, cool core cluster at redshift $z=0.0412$ \citep{oegerle2001}, with an estimated
mass $M_{500} = 1.49\times10^{14} M_{\odot}$ \citep{piffaretti2011}, and a global (i.e., including the cool
core emission) temperature of $\sim 4$ keV within $\sim 0.3r_{500}$ \citep{2006Fujita}. This cluster,
in addition to the historical claims of rotation previously mentioned, is also included among the
clusters with significant rotation in the recent study of \citet[][]{2017Manolopoulou}, and confirmed by
the recent study of \citet{2018Song}, who found a $3.8\sigma$ signal of rotation based on the analysis
of 285 member galaxies within $R<20^\prime$. The corresponding rotation velocity is 380--440 km/s
at $20^\prime$, therefore slightly below the best spectral resolution expected for CCD data.
Despite this, we aim at exploiting the archival {\sl Chandra} observation of A2107 to search for global
ICM rotation and compare it to the optical results.

The paper is organized as follows. In Section 2, we briefly describe the expected signatures of ICM rotation
in the X-ray band in the simplest, idealized cases.  In Section 3, we describe data reduction and analysis.
In Section 4, we describe the measurement of global rotation, and present the results.  In Section 5,
after commenting on the constraints of possible systematics in our measurements,
we discuss the rotation versus merger scenario on the basis of our findings and mention possible extensions
of our study. Finally, our conclusions are summarized in Section 6.  Throughout the paper, we adopt
the concordance $\Lambda$CDM cosmology with $\Omega_{\Lambda} =0.7$, $\Omega_m =0.3$, and $H_0 = 70$
km s$^{-1}$ Mpc$^{-1}$. We note that our results have a negligible dependence on the adopted cosmology. Quoted error bars correspond to a 1 $\sigma$ confidence level, unless
noted otherwise.

\section{ICM rotation: expected signatures in X-rays}

\begin{figure*}
\begin{center}
\includegraphics[width=0.33\textwidth, trim=142 220 160 187, clip]{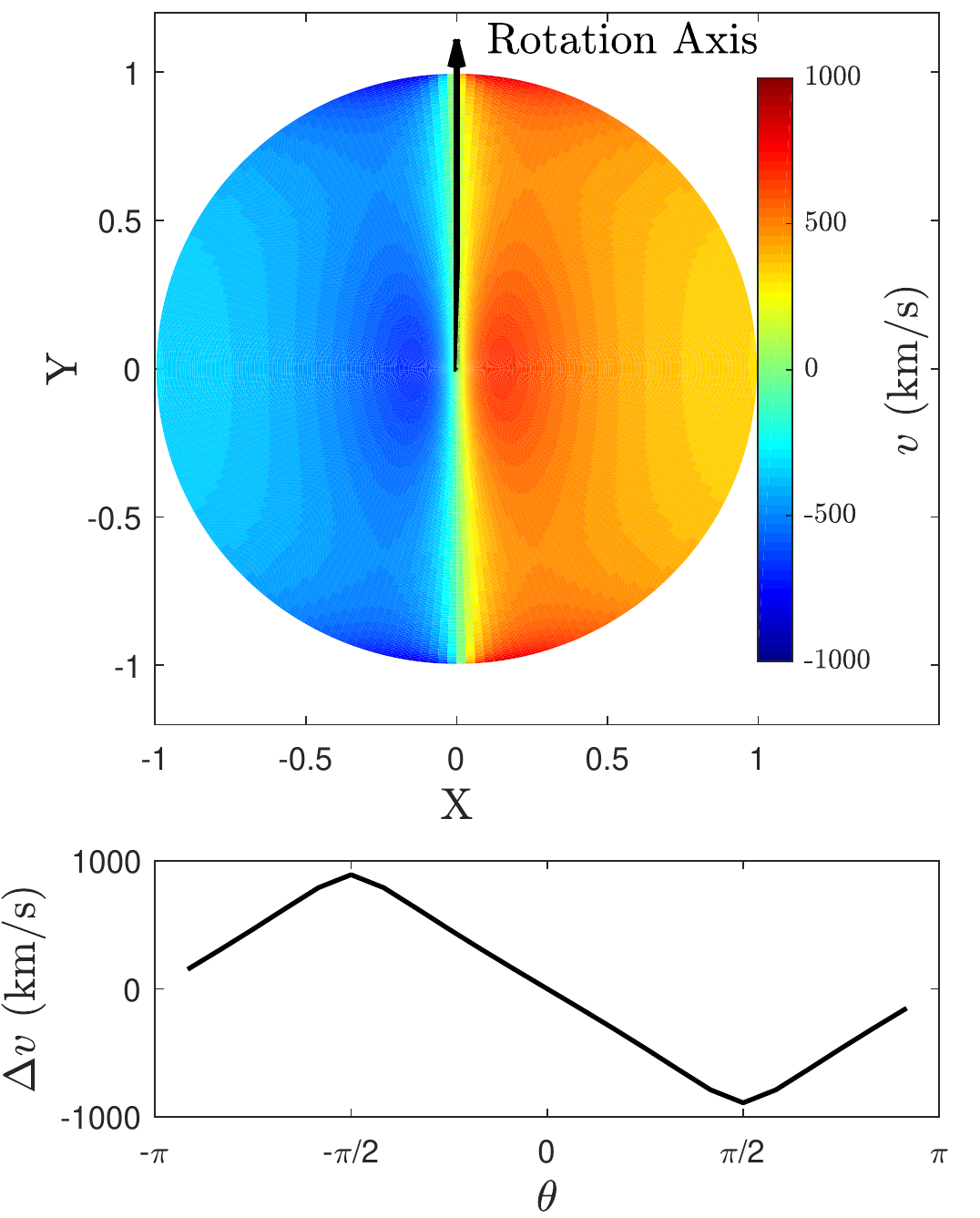}
\includegraphics[width=0.33\textwidth, trim=142 220 160 187, clip]{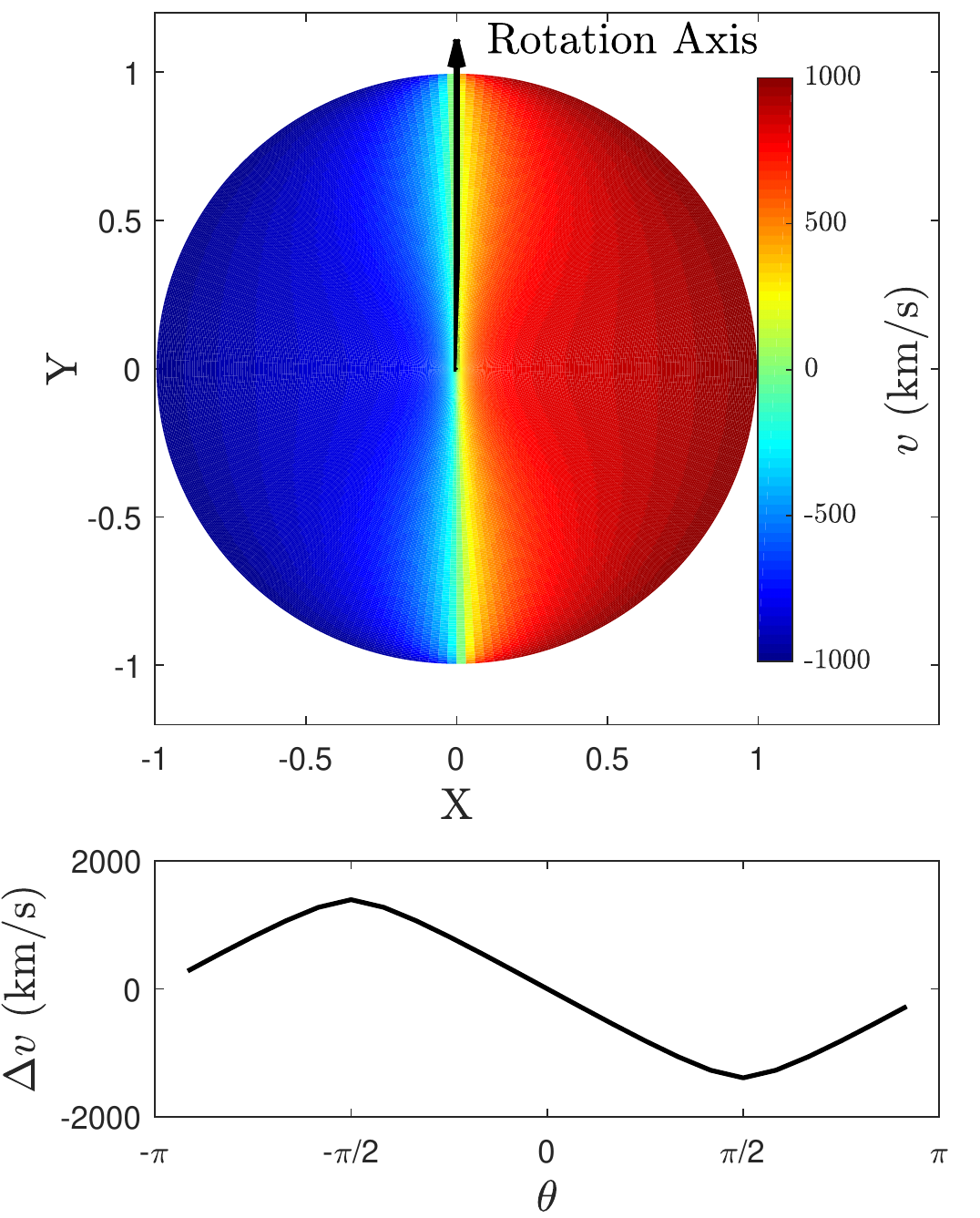}
\includegraphics[width=0.33\textwidth, trim=142 220 160 187, clip]{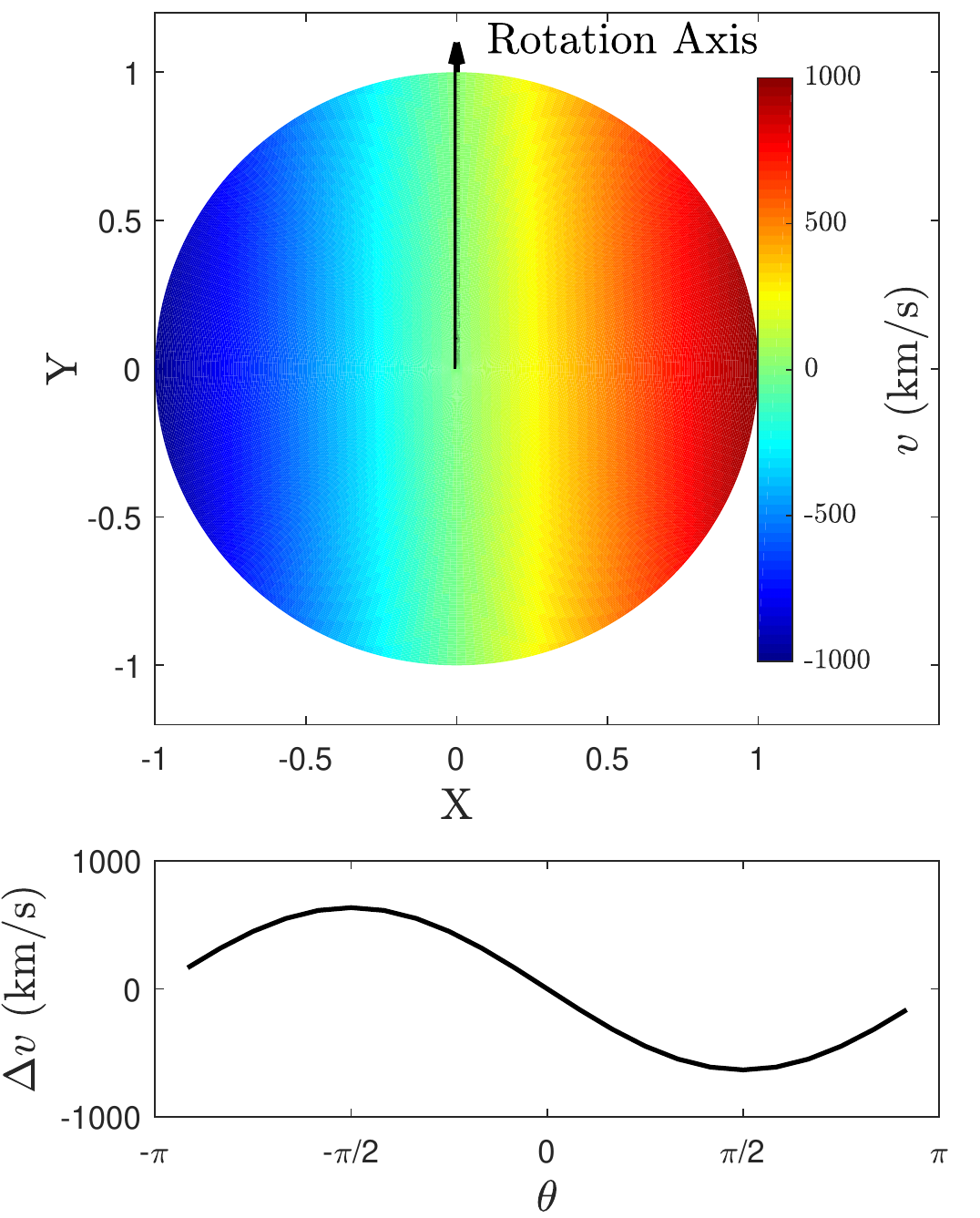}
\caption{Left panel: projected velocity map observed in a cluster rotating with a differential
velocity curve according to equation \ref{rotcurve}, where $c=5$, $c_{\rm v}=10$,
and ${\tt v}_{\rm max}=1000$ km/s, where ${\tt v}_{\rm max}={\tt v}(\lambda_0)={\tt v}_0$/4 as
follows from Equation \ref{eq1}.
Middle and right panels: same as the left panel, but for rotation with a constant velocity of 1000 km/s,
and a rigid body rotation with $\Omega$=1000km/s/$r_{\rm vir}$, respectively. In the lower panels,
we plot the average velocity difference between the two semicircles defined by the dividing line,
as a function of the angle $\theta$ ($\theta=\pi/2$ when the dividing line overlaps with the rotation axis).
}
\label{example}
\end{center}
\end{figure*}

In this section we briefly discuss the expected signature of ICM rotation in the X-ray band,
assuming cylindrical rotation with no dependence along the rotation axis.  We assume a
dependence of the rotation velocity on the distance $\lambda$ from the rotation axis  of the kind

\begin{equation}
{\tt v}(\lambda) = {\tt v}_0 {{\lambda/\lambda_0}\over {(1+\lambda/\lambda_0)^2}}\, .
\label{eq1}
\end{equation}

\noindent
This rotation curve is introduced in \citet{bianconi2013} as representative of a set of plausible rotation
patterns in the ICM, as opposed to simpler but less physically motivated cases like rigid-body
rotation (constant angular velocity) or flat rotation curve (a steep rise followed by a constant velocity).
Here we do not make any attempt to connect the assumed rotation to the hydrostatic distribution
of the ICM, which clearly would depend on the rotation pattern, and on the stability of the ICM
that can be affected by turbulence due to a strong radial gradient in the tangential velocity. Therefore,
we naively consider rotation in an almost spherically symmetric ICM distribution, which is unphysical.
However, this is a good approximation to the case of A2107, which shows a regular morphology with
small eccentricity, as we will discuss in Section 5.

To compute the map of the velocity along the line of sight, we need to include the effects of projection
in the optically thin ICM distribution, and convolve the emission at each position with the
corresponding emission weight.
Clearly, the presence of different velocities projected along each line of sight results in both a
broadening of the line and a shift of the line centroid. The first effect \citep[see][]{2012Zhuravleva}
is not discussed here,
since it is below the capability of CCD spectra, while the shift of the line centroid is potentially
detectable, at least in the most prominent 6.7--6.9 keV emission line complex from the $K_\alpha$ lines
of hydrogen- and helium-like iron.  To compute the emission-weighted shift at each position on the plane of
the sky, we need to assume a 3D density distribution of the ICM.  The average ICM velocity (or redshift shift
with respect to the global cluster redshift) along each line of sight is thus weighted by the
emissivity, which, in turn, is given by the electron density squared $n_{\rm e}^2$
multiplied by the cooling function $\Lambda(T,Z)$.
The dependence of the emission weight on temperature and metallicity is relevant when both quantities
rapidly vary with the radius, which typically occurs in cool cores.  For simplicity, here we consider only
the case of a smooth, isothermal ICM distribution with uniform metallicity described by a
single $\beta$-model \citep{1976Cavaliere}, with an emission weight proportional simply to $n_{\rm e}^2$.

Each line of sight sees the contribution of different rotation velocities corresponding to
different distances from the rotation axis, and each contribution is weighted by $n_{\rm e}^2$.
Therefore, the resulting projected velocity map depends on the ratio of the two scale length:
the core radius $r_c$ of the $\beta$-model distribution of the ICM, and the scale length $\lambda_0$ of the
velocity curve. The detailed derivation of the projected line of sight velocity map
is provided in the Appendix.  An example with the rotation axis lying in the plane of the sky,
is shown in Figure \ref{example}, where all the quantities
are expressed in terms of the virial radius $r_{\rm vir}$. The relevant parameters are therefore the
concentration of the ICM 3D density distribution, defined as $c=r_{\rm vir}/r_{\rm c}$,
and the concentration of the velocity curve, defined as $c_{\rm v}=r_{\rm vir}/\lambda_0$. In the left panel of
Figure \ref{example}, the $\beta$ parameter is set to 2/3, while the concentration parameters $c$ and
$c_{\rm v}$
are put equal to 5 and 10, respectively.  For example, this implies, for a virial radius of 1.5 Mpc, a
core radius $r_{\rm c}=300$ kpc, and a velocity scale of $\lambda_0=150$ kpc, similar to the model adopted
in \citet[][]{bianconi2013}. For the sake of comparison, we show also the projected map in the case of
a flat rotation and a rigid-body rotation (central and right panel, respectively).  In all the cases the
maximum velocity is set to ${\tt v}_{\rm max}=1000$ km/s. In the case of a rotation profile
described by Equation \ref{eq1}, the velocity
reaches its maximum for $\lambda=\lambda_0$, and corresponds to ${\tt v}_{\rm max}={\tt v}_0/4$.

In the left panel of Figure \ref{example}, the projected velocity map shows a peculiar shape that
reflects the combined effects of the spherical ICM distribution and the cylindrical rotation pattern.
This pattern can be, in principle, identified in high resolution redshift maps. However, given the poor
spectral resolution and the need to extract spectra from large regions to increase the S/N,
we can simply assume a constant projected velocity along lines parallel to the rotation axis.
The rotation axis can be efficiently identified from simple observables, such as
the average emission-weighted projected velocity in opposing semicircles.
In the lower panels of Figure \ref{example} we show the difference in the projected
velocity averaged in each half of the cluster as a function of the angle $\theta$ (defined
as in Section 4, with $\theta=\pi/2$ when the dividing line overlaps with the rotation axis).

From these examples we conclude that a simple analysis of CCD spectra with high S/N
can identify the rotation axis, the maximum velocity and the scale length of the rotation curve,
as long as the maximum velocity difference is comparable or larger than the spectral resolution of CCD data.
This simple modelization does not take into account the flattening of the ICM distribution
due to the rotation itself \citep[see][]{bianconi2013}, nor, obviously, the presence of disordered
bulk motions across the ICM, which can in principle overlap with the rotation pattern.
Therefore, in the following we assume that rotation, if present, occurs around an axis in the plane of
the sky, and that there are no significant bulk motions.  We will consequently proceed in two steps:
first, we identify the rotation axis by maximizing the velocity difference between semicircles as a
function of the angle of the dividing line; then, we measure the velocity in stripes parallel to the preferred rotation axis previously identified.  Finally,
we fit the observed rotation curve with Equation \ref{eq1} to derive the best-fit values of
${\tt v}_0$ and $\lambda_0$.

\section{Data reduction and analysis}

\begin{figure}
\begin{center}
\includegraphics[width=0.49\textwidth, trim=122 228 128 217, clip]{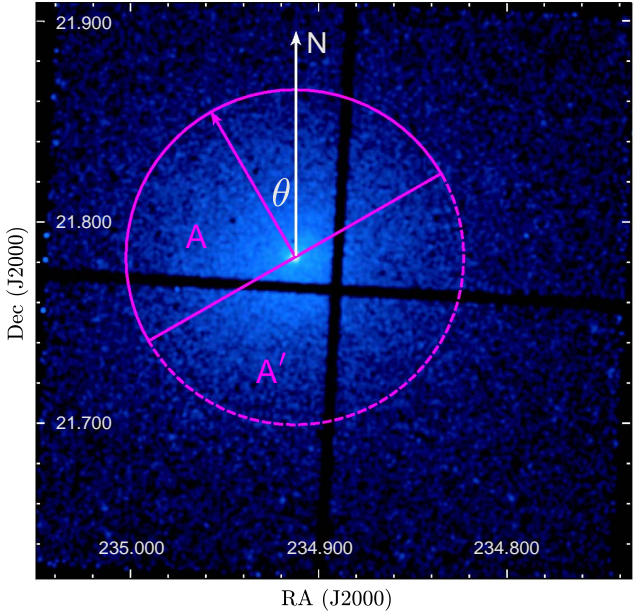}
\caption{{\sl Chandra} X-ray image of A2107 in the 0.5--7~keV energy range. We show the extraction regions
and the definition of the position angle $\theta$ used to measure the projected velocity difference
between semicircles and identify the preferred rotation axis.
In the example shown in the figure, the axis (magenta arrow) with an orientation $\theta=30^\circ$
divides the cluster into the two semicircles A and A$^\prime$.
}
\label{image}
\end{center}
\end{figure}

A2107 was observed with {\sl Chandra} ACIS-I in VFAINT mode on September 7, 2004 for a total
exposure time of 35.4 ks (ObsID 4960). The data are reduced with {\tt CIAO} 4.10,
using the most updated {\sl Chandra} Calibration Database at the time of writing (CALDB 4.7.8).
Unresolved sources within the ICM are identified with {\tt wavdetect},
checked visually and eventually removed. The cluster appears as a strong, extended source
centered on the aimpoint, and covering a significant fraction of each of the four CCDs of ACIS-I
(see Figure \ref{image}).
The extraction radius we use to search for rotation is $\sim 5^\prime$, corresponding to $\sim$240~kpc.
This roughly corresponds to $\sim 0.3 r_{500}$, where the radius $r_{500} = 0.796$ Mpc has
been estimated in \citet{piffaretti2011} using the scaling relation between luminosity $L_{500}$
and mass $M_{500}$.  This radius has been chosen in order to maximize the
S/N, and it includes a total of  $6\times10^4$ net counts in the 0.5--7 keV band. Since the ICM emission is
spread across the entire ACIS-I field of view, the background spectrum is extracted from the
`blank sky' files with the {\tt blanksky} script.

The X-ray spectral analysis is performed with {\tt Xspec} v12.9.1 \citep{arnaud1996}. A double
{\tt apec} thermal plasma emission model \citep{smith2001} is used to fit the ICM spectra.
Galactic absorption is described by the model {\tt tbabs} \citep{2000Wilms}, where the
Galactic HI column density is set to the value measured in \citet{2005Kalberla}, which, at the
position of A2107, is $n_\textrm{H}= 4.46\times 10^{20}$ cm$^{-2}$.  During the fit,
the temperature, metallicity and normalization of the two thermal components are left
free, while the redshift is unique for both.  The use of a double {\tt apec} model is required to avoid
possible bias associated with the presence of multiple temperatures, which may reflect in a slight
but relevant change of the centroid of the line emission complex if the spectral model is forced
to have a single temperature.  We also
check that the use of the 2--7 keV band, as opposed to the full 0.5--7 keV band, gives
consistent results on the redshift. Finally, since the measurement of $n_{\textrm H}$ often has large uncertain and low spatial resolution, for example \citet{starling2013} reports $n_{\textrm H}=5.35\times 10^{20}$cm$^{-2}$ at the position of A2107, we also repeated the fits leaving the value
of $n_\textrm{H}$ free to vary, and find values ranging from $5.0$ to $11.0 \times 10^{20}$ cm$^{-2}$
when analyzing spectra extracted in annular bins at different radii.  We do not consider this as a
reliable indication for a larger value for $n_\textrm{H}$, mostly because the use of two {\tt apec}
models at different temperature increases the degeneracy with the Galactic absorption. Our main concern here
is that leaving free the $n_\textrm{H}$ parameter does not change significantly the best-fit redshift.  To
summarize, this strategy allows us to keep under control the effects of colder gas
and the complex interplay of different metallicity values at different temperatures, following the
prescription discussed in \citet{liu2015}.

\begin{table}
\centering
\caption{The {\sl Chandra} data used in this work. Redshift values refer to optical estimates from the 
NASA/IPAC Extragalactic Database (NED). In the fourth column we list the total exposure times 
after data reduction. }
\begin{tabular}{lccc}
\hline
Cluster   & $z$ & {\sl Chandra} ObsID & Exptime (ks)   \\
\hline
A2107 & 0.041 & 4960 & 35.4 \\
A2029 & 0.077 & 4977 & 77.1 \\
A1689 & 0.183 & 5004,6930,7289,7701 & 175.9 \\
\hline
\end{tabular}
\label{data}
\end{table}

We also reduce the data of other two clusters observed with {\sl Chandra},
selected in order to have a comparable data quality: A2029 and A1689.  These two targets do not have
reported claims of rotation, and are considered relaxed clusters with regular morphology.
We use these two targets as a basic control sample to test our strategy
in searching and characterizing global rotation.  In particular, we check whether a spurious rotation
signature may appear as a result of our assumption.  Clearly, a much larger
control sample would be required to assess the statistical significance of a particular rotation measurement.
This goes beyond the goal of this study.  All the {\sl Chandra} data used in this work, the ObsID
and the total exposure time after data reduction are listed in Table \ref{data}.

\section{Measurement of global rotation}

To identify and quantify global rotation in our sources, we follow the two-step strategy
described in Section 2.  First, we identify the preferred axis of rotation, defined as the projected
line that maximizes the velocity difference measured between the two semicircles.
In practice, as shown in Figure \ref{image} we divide the cluster into two semicircles
$A$ and $A^{\prime}$, with orientations $\theta$ and $\theta-\pi$
where $\theta$ is the angle from north counterclockwise to the vertical axis of semicircle $A$.
We perform our spectral analysis of the total emission extracted from the regions
$A$ and $A^{\prime}$, and, in particular, we measure the best-fit redshift. This value
corresponds to the emission-weighted average redshift of all the ICM components observed along
the line of sight in the selected region.

\begin{figure}
\begin{center}
\includegraphics[width=0.49\textwidth, trim=80 215 100 230, clip]{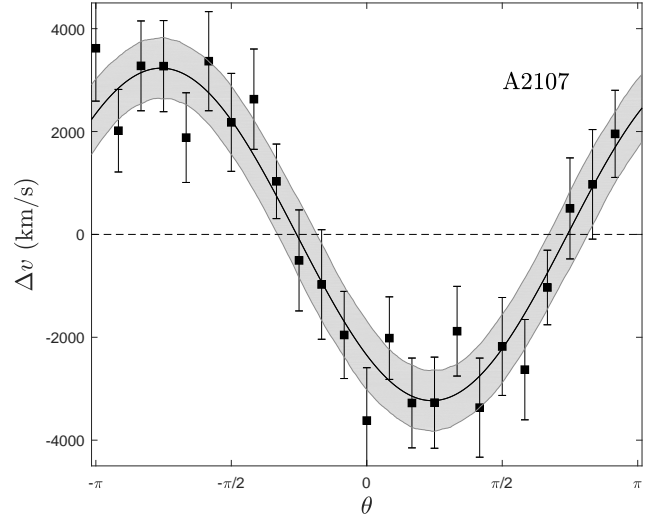}
\caption{ The $\Delta {\tt v}-\theta$ plot of A2107 measured in steps of $\Delta \theta = 15^\circ$.
The angle $\theta$ ranges from -180$^\circ$ to 180$^\circ$, increasing counterclockwise from north
($\theta=0^\circ$) to east ($\theta=90^\circ$), and decreasing clockwise from north to west
($\theta=-90^\circ$).  Error bars correspond to 1 $\sigma$ uncertainty.
}
\label{rotation}
\end{center}
\end{figure}

The measurement of ICM redshift and the assessment of its uncertainty have been described and applied
in \citet{liu2015,liu2016} and \citet{liu2018b}.  From the difference of the two redshifts, we obtain
the velocity difference between the two regions as $\Delta {\tt v}={\tt v}_{A}-{\tt v}_{A^{\prime}}=c \times
(z_{A}-z_{A^\prime})/(1+z_{\rm cl})$, where $z_{\rm cl}$ is the cluster redshift $z=0.0412$
\citep{oegerle2001}.  We repeat the measurement
as a function of $\theta$ and sample $\Delta {\tt v}$ as a function of $\theta$ in the range $-\pi <\theta<\pi$.
The $\Delta {\tt v}-\theta$ plot is therefore made of points that are highly correlated, since nearby points
come from spectral fits of overlapping regions. The resulting curve is then fitted with the function
$\Delta {\tt v} = \Delta {\tt v}_{\rm max}\cdot {\rm cos}(\theta-\theta_{\rm max})$, where
$\theta_{\rm max}$ corresponds to the maximum velocity difference. The $\Delta {\tt v}-\theta$ plot
and the best fit function are shown in Figure \ref{rotation}.  The best fit parameters of the curve,
obtained by a simple $\chi^2$ fit, are $\Delta {\tt v}_{\rm max} = 3230\pm 590$ km/s, and
$\theta_{\rm max} = -137\pm 13^\circ$, where the 1 $\sigma$ error bars are obtained by marginalizing
with respect to the other parameter. This result by itself is not a probe of rotation, since a periodic
curve is always obtained in any cluster with this method, simply because of noise, or because of the
presence of some bulk motion in a particular region of the cluster.  Clearly, a value of
$\Delta {\tt v}_{\rm max}$ significantly higher than zero constitutes a strong indication
that the dynamical properties of the ICM are far from being consistent with the usual hydrostatic,
no-rotation scenario.

\begin{figure}
\begin{center}
\includegraphics[width=0.49\textwidth, trim=80 215 100 230, clip]{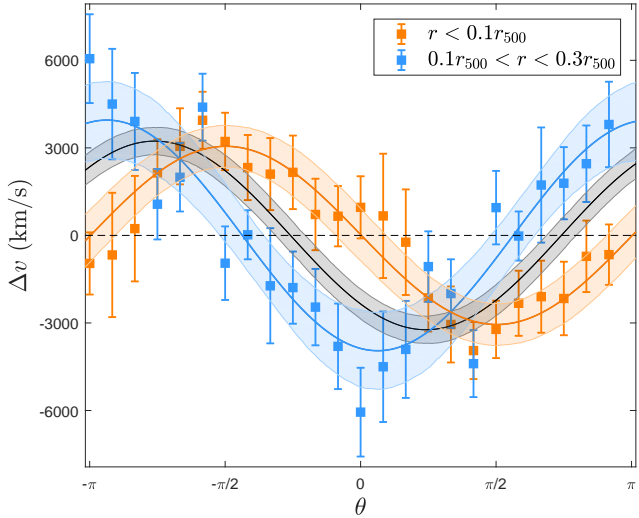}
\caption{The velocity difference in A2107 measured across the rotation axis identified by $\theta$
at two different radial ranges. The black line shows the best-fit curve in Figure \ref{rotation}.
 }
\label{rotation3}
\end{center}
\end{figure}

A large uncertainty on the rotation axis may suggest that the shape of the curve is not in agreement
with that expected from coherent rotation.  To investigate this aspect, we repeat the same measurement
using semi-annuli rather than semicircles, in order to search for rotation in shells.
In Figure \ref{rotation3} we show the rotation curves obtained with the same method but
in different radial ranges, namely $r < 0.1r_{500}$ and $0.1r_{500} < r < 0.3r_{500}$. The best fit parameters of the rotation curves corresponding to different radii are listed
in Table \ref{para}.  We find that the two measurements of $\theta_{\rm max}$ obtained at
different radii are inconsistent with each other at more than 3 $\sigma$, showing that a complex,
non-cylindrical rotation pattern may be more adequate to describe the results.   This may well be a
hint for the presence of asymmetric bulk motions, however the data quality is not high enough to further
investigate this possibility. Therefore, in the following we will assume
$\theta_{\rm max}^{\rm ICM}=-137\pm 13^\circ$.

\begin{table}
\centering
\caption{The best fit parameters of the rotation curves corresponding to different radii. }
\begin{tabular}{cccc}
\hline
Radius   & $\Delta {\tt v}_{\rm max}$ (km/s) & $\theta_{\rm max}$   \\
\hline
0--0.3$r_{500}$ & $3230\pm590$ & $-137\pm13^{\circ}$ \\
0--0.1$r_{500}$ & $3050\pm680$ & $-99\pm16^{\circ}$ \\
0.1--0.3$r_{500}$ & $3950\pm1170$ & $-169\pm16^{\circ}$ \\
\hline
\end{tabular}
\label{para}
\end{table}

After identifying the preferred rotation axis of the ICM and its uncertainty, we build the rotation curve
by slicing the ICM into a number of independent linear regions parallel to the rotation axis.  The
average projected redshift is measured in each of these slices, whose size is chosen in order to
have at least $10^4$ net counts in the 0.5--7 keV band in each, and therefore a reasonable statistical
error on the redshift. We identify four regions, where we compute the velocity difference with respect to
the global redshift of the ICM, which is measured to be $z_{X} = 0.0393\pm 0.0014$ by fitting the spectrum
of the global emission within 5$^\prime$. Incidentally, we note that the X-ray redshift is
slightly more than 1$\sigma$ lower than the optical redshift value $z_o=0.0412$. This difference is
not statistically significant, however it may be the hint of a disturbed ICM dynamics, possibly decoupled
from the dynamics of the collisionless mass components (galaxies and dark matter), as may happen
during mergers \citep[see][]{liu2018b}. Moreover, the X-ray redshift we measure here is based on a smaller
region of $5^\prime$ radius, with respect to the $20^\prime$ radius considered by \citet{oegerle1992} 
and \citet{oegerle2001} to compute the average optical redshift.

\begin{figure}
\begin{center}
\includegraphics[width=0.49\textwidth, trim=80 215 100 230, clip]{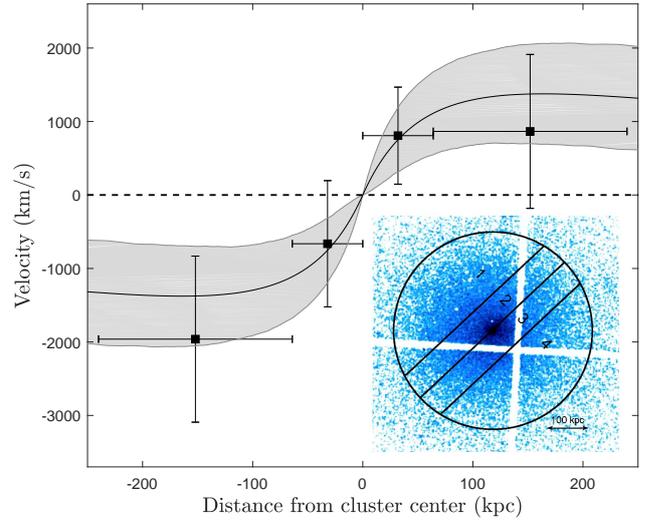}
\caption{Velocity gradient across the slices parallel to the rotation axis defined in Figure \ref{rotation3}.
The solid line shows the best-fit function and the shaded area shows the 1 $\sigma$ confidence interval.
}
\label{slice}
\end{center}
\end{figure}

We can now compute the velocity of each slice as ${\tt v}_i = c\times (z_i-z_{X})/(1+z_{X})$
where $i$ is the index of the slice.  The result is shown in Figure \ref{slice}. If we fit these four
points with the same rotation curve described in Section 2,
we find ${\tt v}_0 = 5500\pm 2400$ km/s, corresponding to ${\tt v_{\rm max}} = 1380 \pm 600$,
and $\lambda_0 = 165\pm 110$ kpc. The $\chi^2$ value of the fit is below unity because of the large
error bars ($\chi^2=0.48$), while the $\chi^2$ value for no rotation is 5.79, which corresponds to
rejection at a 95\% level for 3 degrees of freedom.

\begin{figure*}
\begin{center}
\includegraphics[width=0.49\textwidth, trim=80 215 100 230, clip]{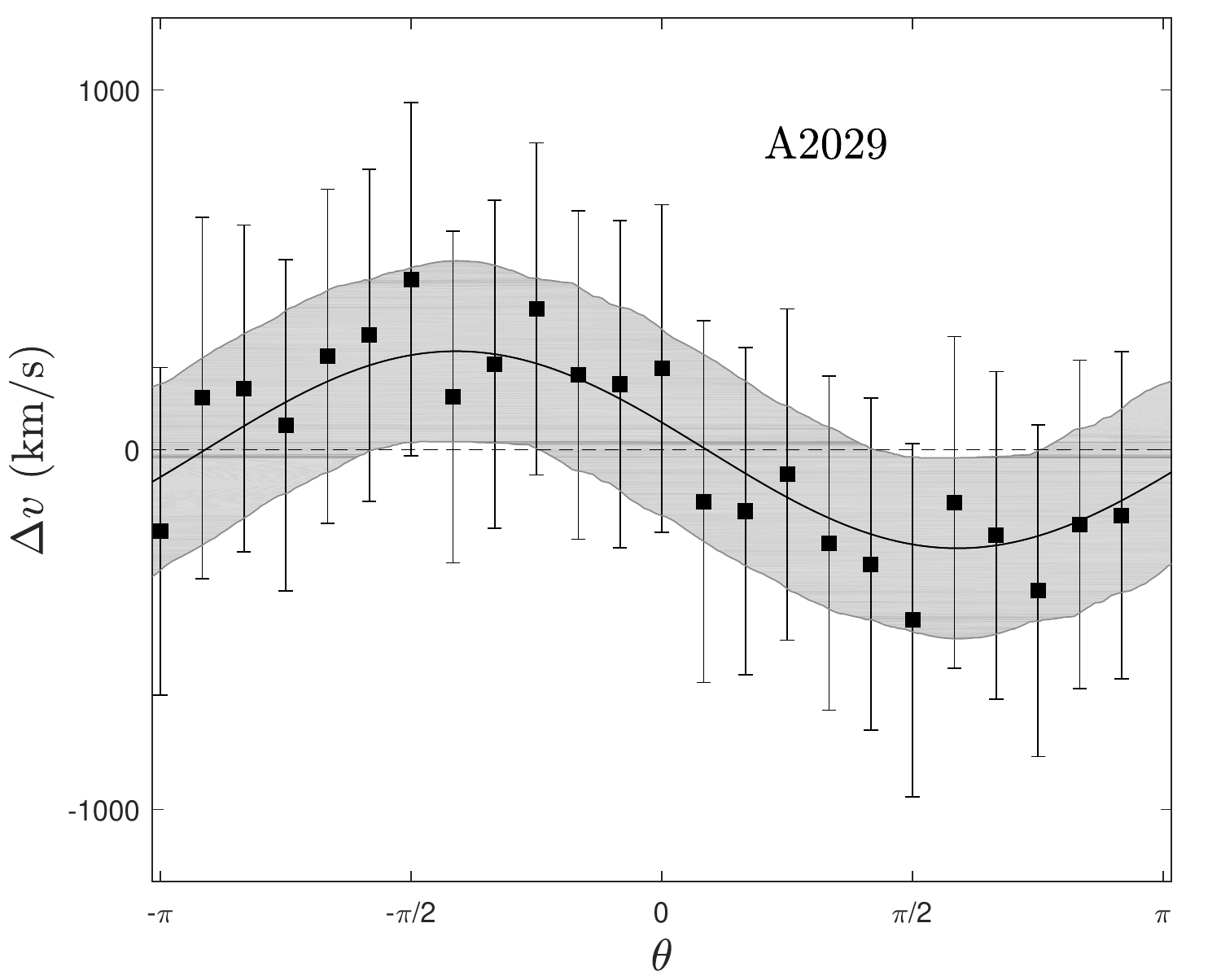}
\includegraphics[width=0.49\textwidth, trim=80 215 100 230, clip]{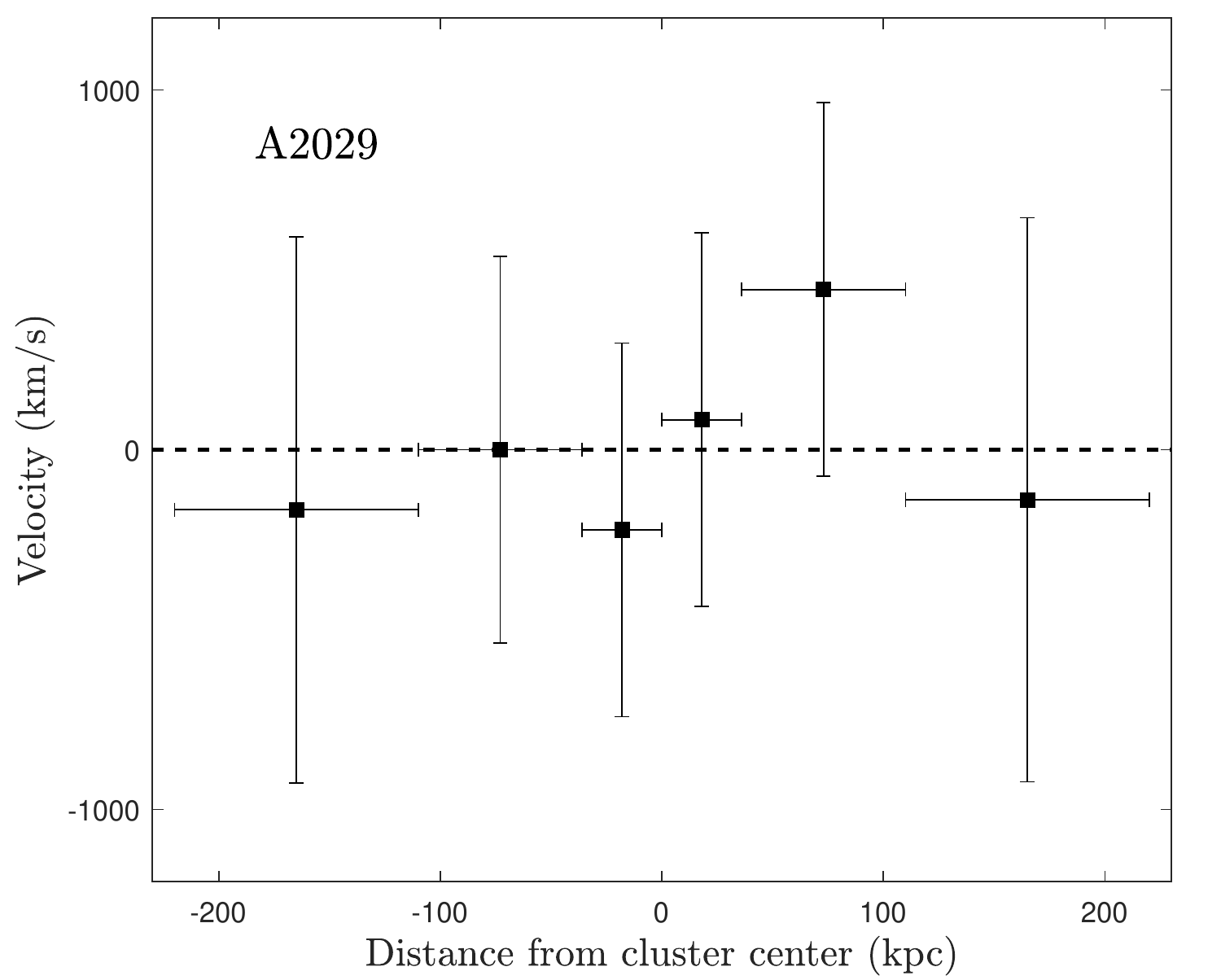}
\includegraphics[width=0.49\textwidth, trim=80 215 100 230, clip]{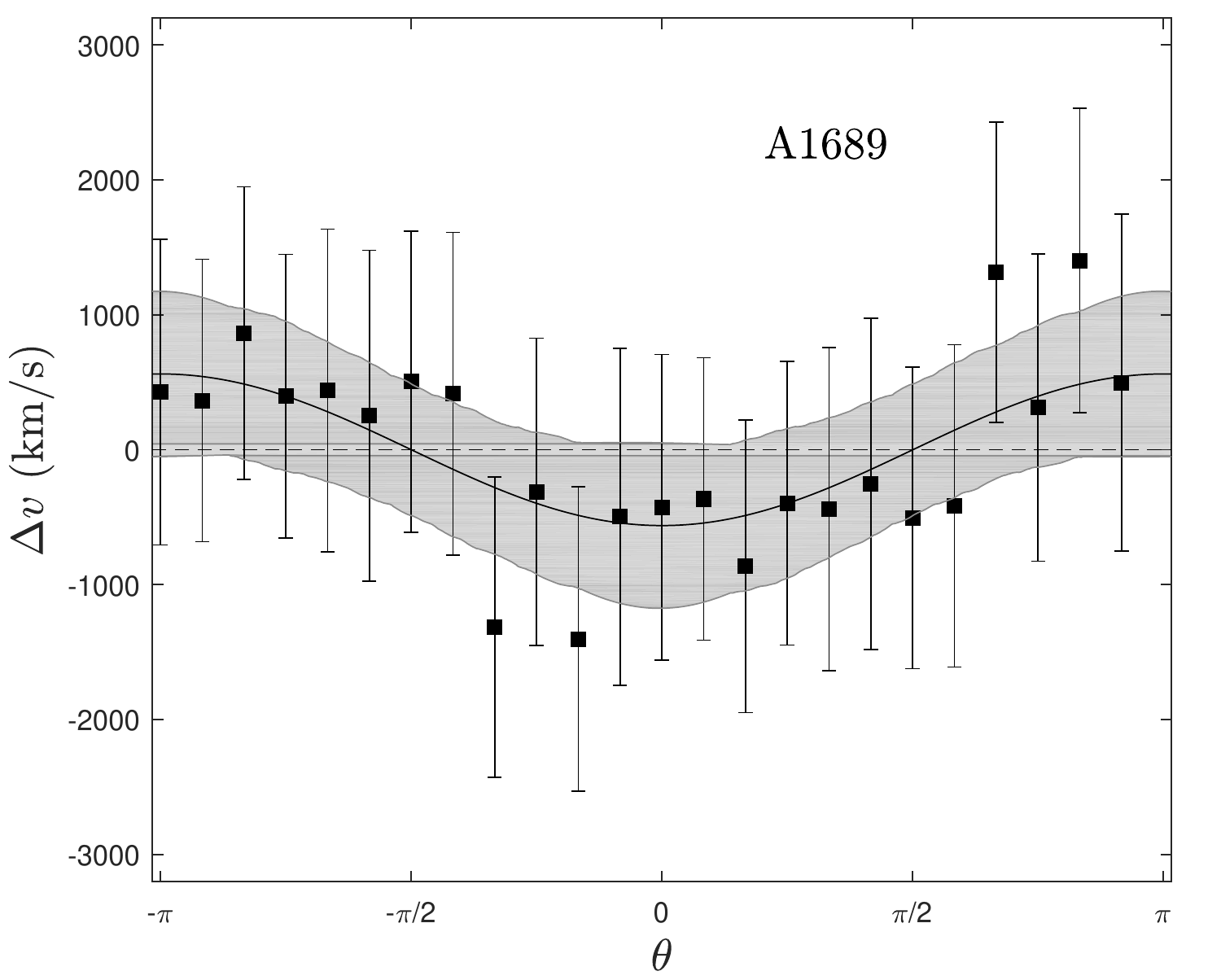}
\includegraphics[width=0.49\textwidth, trim=80 215 100 230, clip]{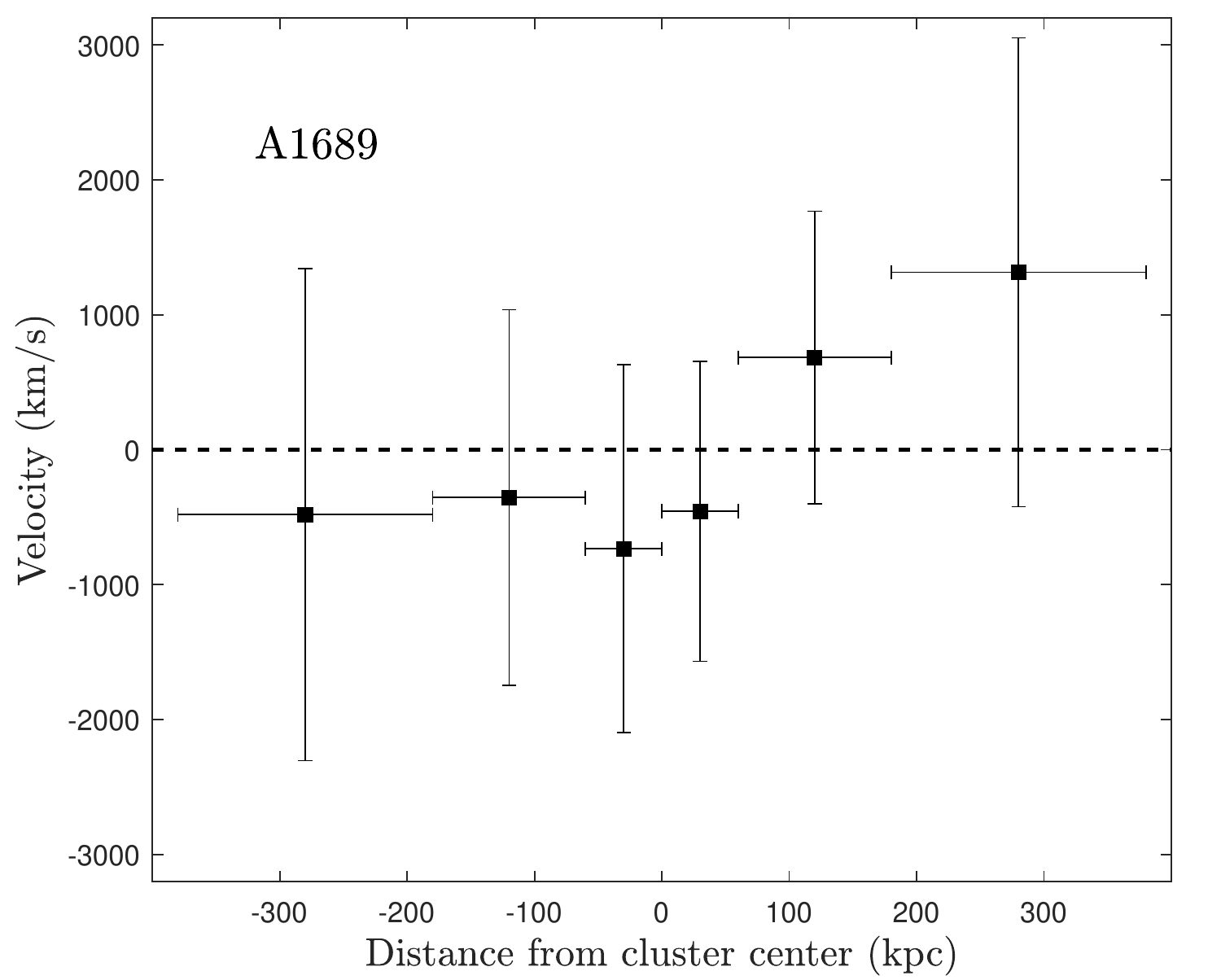}
\caption{Results for the control sample A2029 and A1689. {\sl Left}: $\Delta {\tt v}$-$\theta$ curves.
{\sl Right}: Velocity gradient along the slices parallel to the rotation axis defined by the
$\Delta {\tt v}$-$\theta$ curve. }
\label{rotation2}
\end{center}
\end{figure*}

We repeat the same analysis on the two relaxed clusters A2029 and A1689, where no significant rotation
is expected.  In Figure \ref{rotation2}, we show the $\Delta {\tt v}-\theta$
curves of these two clusters, where we do not find signatures of rotation.
As already noticed, formally a rotation axis
can be found for both clusters, but with large uncertainty, and on the basis of a non-significant
maximum redshift difference. We proceed with the measurement of the rotation curve, slicing the
clusters parallel to the rotation axis. The rotation curves are shown in the right panel
of Figure \ref{rotation2}.
Both profiles are consistent with no rotation. Assuming no rotation in these two clusters
results in $\chi^2$ of 1.02 and 1.56 with 5 degrees of freedom.
Clearly, we cannot exclude the presence of rotation in these clusters on the basis of our data,
but we exclude a rotation signature with the same significance found for A2107.

We conclude that our spectral analysis of the ICM is consistent with a rotation reaching a maximum velocity
of 1000--2000 km/s at a radius between 100 and 300 kpc.  The velocity difference
between two sides of the cluster is significant at more than $2\sigma$. We also find a null result in two
regular clusters, showing that there are no obvious systematic effects in our analysis strategy
that can mimic the presence of rotation, providing thus support to the capability of our analysis in finding
velocity difference across the ICM.

\section{Discussion}

Before discussing the implications of our findings, we rapidly review possible systematics
that may affect our results.  Uncertainties on the best-fit redshift values due to calibration issues
and fluctuation of the gain of CCDs as a function of the epoch of observation have been discussed in
detail in \citet{liu2015} and \citet{liu2016}. In this work, we only use a single observation,
so the time variation of CCD is negligible.  Calibration may also change as a function of the position on the
CCD.  We perform a quick check by computing the position of the Au L$\alpha$ and Ni K$\alpha$
fluorescent lines in the spectra obtained from the extraction regions used in our analysis and from the
corners of the CCD.  Within the statistical limits due to the modest exposure time we are not able to find
any significant difference in the centroid of the lines in difference places.  Despite we are not able
to firmly rule out the presence of some gain variation across the CCD, we are nevertheless able
to estimate its impact to be less than the current statistical error on our redshift measurements.
We also refer to the extensive investigation on calibration issues on redshift measurement with {\sl Chandra}
presented in \citet{liu2015}, where we constrained these effects to be at maximum a 5--10\% of the typical
statistical error, and therefore not relevant to our conclusions.

We also explore the impact of the uncertainty of background modelization on the redshift measurement.
Because the background spectrum we used is generated from the {\sl Chandra} `blank-sky' files, its
normalization in the hard energy range may not be appropriate for our observation.  We conservatively
consider a 10\% maximum variation in the normalization of the background spectrum, and verify that this
reflects in a fluctuation in redshift of the order of $\sim$5\%, which is well below the statistical
error, confirming that the redshift gradient shown in Figure \ref{slice} is robust against
uncertainties in the background modelization.

\begin{figure}
\begin{center}
\includegraphics[width=0.49\textwidth, trim=130 260 148 263, clip]{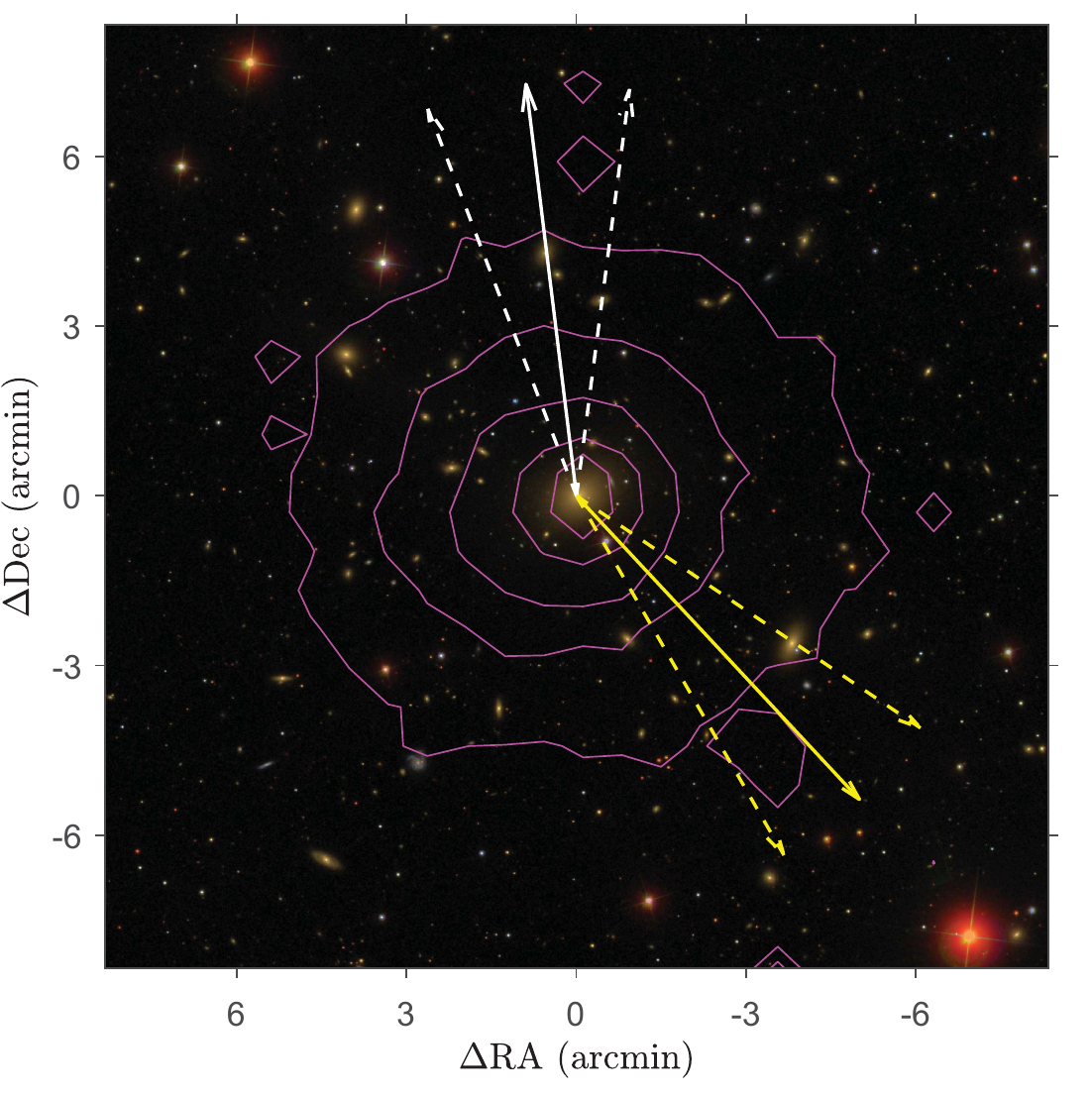}
\caption{SDSS RGB image of A2107 obtained from $g,r,i$ band images.  Magenta contours correspond
to isophotes in the {\sl Chandra} X-ray image. White and yellow arrows mark the preferred direction of
rotation of the member galaxies within 0--20$^\prime$ \citep[from][]{2018Song} and of the ICM, respectively.
Dashed arrows show the 1 $\sigma$ uncertainty on the angle of the rotation axis.}
\label{sdss}
\end{center}
\end{figure}

We then compare our results with the rotation identified by the spectroscopy of the member galaxies in
\citet{2018Song}. In this work, the orientations of the maximum velocity are measured in 3
different bins:
$0^\prime\le r < 20^\prime$, $20^\prime\le r < 35^\prime$, and $35^\prime\le r < 60^\prime$.
Since the extraction radius in our X-ray analysis is only 5$^\prime$,
we compare our result with that of the [0$^\prime$--20$^\prime$] optical bin.  We find that
the rotation axis are significantly different, with $\theta_{\rm max}^{\rm gal}=7\pm14^\circ$,
and $\theta_{\rm max}^{\rm ICM}=-137\pm13^\circ$, with the corresponding momentum vectors pointing almost
in opposite directions.  In Figure \ref{sdss}, where the X-ray surface brightness contours from
{\sl Chandra} are overlaied to the optical image of A2107 from the Sloan Digital Sky Survey, we also
show the preferred rotation axis from the optical and X-ray data.
Despite we are comparing rotation velocities estimated at different scales, we
are surprised to find a rotation axis with a direction completely different from the optical one.
It is not surprising to observe the ICM dynamically decoupled from the galaxies, however, in the case
of global rotation, we should have observed a consistent rotation axis.  Incidentally, if we
consider the 20$^\prime$--35$^\prime$ radial bin in \citet{2018Song} with 
$\theta_{\rm max}=-150\pm36^\circ$, we find that the rotation axis of the galaxies and
of the ICM are consistent within the statistical errors.  Clearly, this may well be just a coincidence, 
and our findings seem to suggest that the velocity pattern is probably not uniform, 
and that the galaxies and ICM do not share the same projected velocity across the cluster.

Another critical aspect is the effect on the total mass measurement implied by the temperature
profile and the measured rotation.  The total mass without considering rotation is
computed via the hydrostatic equation:
\begin{equation}
\frac{\nabla P}{\rho} = -\nabla \Phi = -\frac{GM(r)}{r^2},
\label{eq2}
\end{equation}
where $\rho$, $P$ and $\Phi$ denote the gas density, pressure, and the total gravitational
potential, respectively. Inserting $P=nk_{\rm B}T=\rho_{\rm g} k_{\rm B}T/(\mu m_{\rm p})$, the total mass within $r$ is:
\begin{equation}
    M(r) = -\frac{k_{\rm B} T(r) r}{G\mu m_{\rm p}}\left(\frac{d{\rm ln}\rho_{\rm g}(r)}{d{\rm ln}r}
    +\frac{d{\rm ln}T(r)}{d{\rm ln}r}\right).
\label{eq3}
\end{equation}

\begin{figure}
\begin{center}
\includegraphics[width=0.49\textwidth, trim=85 200 95 233, clip]{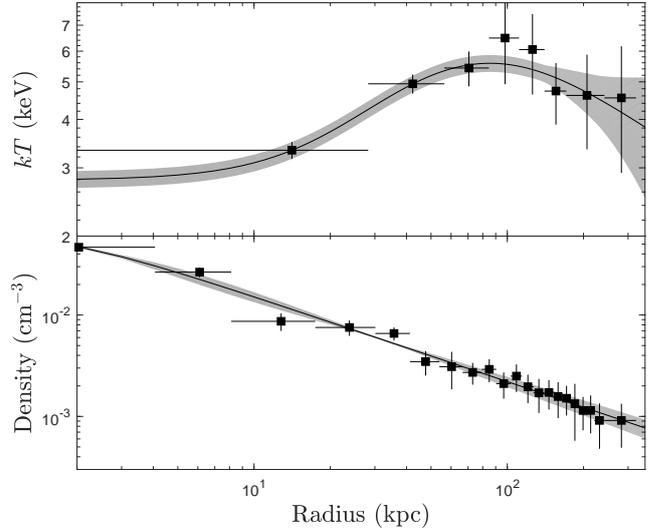}
\caption{Deprojected temperature (upper panel) and density (lower panel) profiles of A2107. }
\label{Tprofile}
\end{center}
\end{figure}

\begin{figure}
\begin{center}
\includegraphics[width=0.49\textwidth, trim=95 200 105 215, clip]{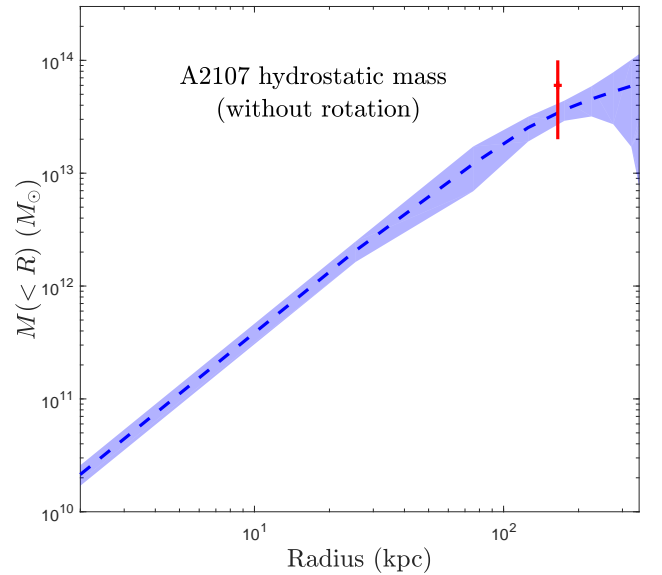}
\caption{The hydrostatic mass profile of A2107 (without including the effect of global rotation) with
1 $\sigma$ uncertainty computed with bootstrapping shown as a shadowed area. The cross
is the approximated mass-correction term due to rotation along the $z=0$ plane at $\lambda_0 = 160$ kpc. }
\label{mass}
\end{center}
\end{figure}

\noindent
We measure the deprojected temperature and density profiles using 8 and 20 bins, respectively, with 
the routine DSDEPROJ \citep[see][]{sanders2007}. We then fit the temperature profile with the model proposed by
\citet{vikhlinin2006a}, and the density profile with a single $\beta$ model.
The deprojected temperature and density profiles are shown in Figure \ref{Tprofile}.  The best fit
functions are used to compute the logarithmic slopes
in Equation \ref{eq3}.  The hydrostatic mass profile is shown in Figure \ref{mass}. With very
large uncertainties in the outskirts, we measure a value of
$(4.5 \pm 1.4)\times 10^{13} \, M_\odot$ at $r=200$ kpc, in agreement with the mass profile
obtained from the galaxy velocity dispersion by \citet{kalinkov2005}.

If we consider hydrostatic equilibrium in presence of rotation, we are assuming that rotational motions
do not affect motions along the radial direction.  Therefore, we should consider a term proportional
to $({\tt v}^2/\lambda)_{\rm rad}$ in equation \ref{eq3}, which is the projection of the centripetal force on the
radial direction at each point.  A self consistent treatment can be found in \citet{bianconi2013}.
Here we do not make any attempt to solve for the ellipsoidal mass distribution implied by the
presence of rotation and hydrostatic equilibrium.  However, we can compute the mass term along the
direction perpendicular to the rotation axis, that, at the radius $\lambda_0$ corresponds to a correction of
$\sim 6 \times 10^{13} M_\odot$.  From Figure \ref{mass} it is possible to verify that this correction term
is comparable, if not larger, to the hydrostatic mass computed in the assumption of no rotation at
the same radius.  This clearly shows that the spherical symmetry is inconsistent with such a large
rotation. On the other hand, given the large uncertainties, if we consider the 2$\sigma$ lower limit
to the mass term due to rotation, we find a correction of the order of a few $10^{12} M_\odot$,
which corresponds to
$\sim 10$\% of the hydrostatic mass at the same radius, in better agreement with the cluster morphology.

The best-fit value of ${\tt v}_{\rm max}$ is therefore at variance with the regular and round isophotes
of the surface brightness of A2107, that can be appreciated in Figure \ref{sdss}.
We fit the X-ray surface brightness distribution with a 2-D elliptical $\beta$ model, and get an
ellipticity of 0.097, indicating an almost round morphology, that, as shown, may be reconciled with the
measured rotation only when assuming values lower by $\sim 2 \sigma$ with respect to the best fit.

If we relax the hypothesis of hydrostatic equilibrium, we can consider a major merger scenario as opposed to
rotation.  This scenario is suggested by the measurement of an exceptionally high velocity of the
BCG relative to the bulk of the galaxy population of $\sim 270$ km/s \citep{oegerle1992}.
For some reason, however, this possible major merger along the line of sight is not observed in the galaxy distribution nor is associated with any disturbed morphology in the X-ray image.  In
addition, the BCG is still perfectly centered on the cool core within a few arcsec, as shown in
Figure \ref{sdss}.
On the other hand, the BCG is not active both in X-ray and in radio, an occurrence that,
particularly in the radio, should be observed in the majority of cool cores \citep{2009Sun}.
A possible solution may be provided by an ongoing major merger along the line of sight in its
early stages, when the ICM is already significantly decoupled from the bulk of the galaxy, including the BCG,
and the cool core is still visible but is not `active' anymore (in the sense that it does not feed the AGN in the
BCG).

To summarize, the results obtained in this work do not allow us to conclude unambiguously
that the ICM in A2107 is rotating to some degree.  In particular, it is almost impossible with CCD data to
distinguish a genuine rotation from disordered and asymmetric bulk motion, or from a major
merger along the line of sight. Increasing the depth of the observation may help in reducing the
statistical error on the measured redshift, an improvement that can be better achieved with CCD data 
from XMM-Newton, which has a significantly larger effective area with respect to {\sl Chandra} and therefore it
performs more efficiently in observations where high angular resolution is not mandatory.
A better redshift map, in principle, can provide an unambiguous view of a rotation pattern. 
In particular, it will be possible to determine whether the rotation signature is due to a smooth pattern 
across the entire ICM distribution, or it is merely an effect of one or more mergers, with a patchy 
redshift map.  However, spectral analysis of CCD data would improve rather slowly with increasing exposures, 
and therefore would require a large investment of observing time. 

Another possible strategy is the observation with gratings at constrained angles. In this
last case, the rotation (or the asymmetric bulk motion) should show up more clearly as a shift in the
emission lines in the direction perpendicular to the dispersion, which should be aligned to the rotation axis.
We remark that A2107 has the advantage of a relatively low temperature, so that several emission lines can be 
visible in the energy range probed by gratings. A real breakthrough in the study of the distribution
of angular momentum at cluster scale, may be achieved in the next future only with the advent of 
X-ray bolometers, or, possibly, with better SZ data \citep[as discussed in][]{2002Cooray}.

\section{Conclusions}

In this work we define a strategy to search for signatures of rotation in CCD data, on the basis of a
simple rotation model.  We report the measurement of a possible rotation in the ICM of the cluster A2107,
obtained through a spatially resolved measurement of the redshift inferred from the centroid of the
iron emission line complex at 6.7 and 6.9 keV in {\sl Chandra} ACIS-I spectra.
We identify a preferred rotation axis, and find a significant velocity gradient compatible with a
rotation pattern with maximum tangential velocity
${{\tt v}_{\rm max}}=1380\pm 600$ km/s at a radius $\lambda_0\sim 160$ kpc.

If confirmed, this would be the first detection of ICM rotation.  Although this work has been stimulated
by the previous claims of rotation in A2107 obtained by \citet{2017Manolopoulou} and \citet{2018Song}
on the basis of optical spectra of the member galaxies, our results differ both in the direction of the
preferred rotation axis, and in the amplitude of the rotation curve. In particular, the high velocity
associated with the ICM rotation in our data would be in conflict with the assumption of hydrostatic
equilibrium and with the morphology of the cluster.
We argue that an unnoticed  off-center major merger along the line of sight can be an alternative
explanation of the dynamical status of A2107. Therefore, our analysis confirms the peculiar
dynamical nature of the otherwise regular cluster A2107, but is not able to provide a definitive
answer to the rotation versus merger scenario.  We argue that a discrimination among these two scenarios
should wait for the next-generation X-ray facilities carrying X-ray bolometers onboard, while some
improvements can still be made with further CCD data and
angle-constrained grating spectra, preferably with XMM-Newton. The measurement of ICM rotation is potentially
relevant for investigation of the distribution of angular momentum at cluster scale, which is still a debated
aspect of the gravitational growth of cosmic structures.  Therefore, any insight
that can be obtained on the basis of current X-ray facilities in the next years, particularly before XRISM,
due to launch in the early 2020s, would be extremely useful to refine the analysis strategy in this field.

\section*{acknowledgements}
We thank the anonymous referee for a detailed and constructive report.
We acknowledge financial contribution from the agreement ASI-INAF n.2017-14-H.O.

\bibliography{a2107}

\appendix

\section{Projected velocity map of the ICM with a generic rotation curve}

To compute the projected velocity map for a generic rotation curve in a spherical ICM, we
need to convolve a cylindrical rotation curve with a spherical distribution of
ICM density.  As noted in Section 2, we do not solve for a self-consistent hydrostatic and rotating ICM
distribution in a fixed dark matter potential well, so we simply assume the case of a spherical
symmetry for the ICM distribution
despite its rotation.  This assumption can be considered a fair description only when the
rotational support is a minor correction to the pressure support at each radius.  Our treatment here
is meant only to give us a guideline on how to design the analysis strategy to recover the rotation curve.

The projected velocity is measured from the redshift of the iron complex emission line, which,
at any position on the sky, is the emission-weighted value of the centroid of the lines emitted
by each ICM component intercepted by the line of sight.
To compute the emission-weighted quantity along the line of sight, we
consider a cylindrical reference system with the z axis pointing towards the observer, while
the rotation axis is one of the two axis on the plane of the sky.  With this choice, the
transformation from spherical coordinate $r$, $\theta$ and $\phi$ (to describe the ICM properties) to
the cylindrical coordinates $\rho$, $\theta$ and $z$ (to describe the projected redshift map) assume the
convenient form $ \rho = r {\rm sin}\phi$, $\theta=\theta$ and $ z=r {\rm cos}\phi$

We consider a generic rotation curve that depends only on the distance from the rotation axis
$\lambda$, and it is characterized by an overall normalization and a scale length $\lambda_0$.
The velocity perpendicular to the vector $\hat \lambda$ is thus ${\tt v}({\tt v}_0,\lambda, \lambda_0)$.

The distance from the rotation axis reads
\begin{equation}
\lambda=\sqrt{(\rho\times {\rm cos}\theta)^2+z^2}\, .
\end{equation}

\noindent
The velocity projected along the line of sight is therefore
${\tt v}_{\rm los}={\tt v}({\tt v}_0,\lambda, \lambda_0) \times {\rm cos} \alpha$,
where $\alpha$ is the angle between the velocity vector
and the line of sight.  We can show that we also have ${\rm cos} \alpha = \rho \, {\rm cos}\theta/\lambda$.
If we use a generic weighting function $W(\rho,\theta,z)$, we can express the observed
${\tt v}_{\rm los}$ as

\begin{equation}
{\tt v}_{\rm los}(\rho,\theta) = {{\int  W(\rho,\theta,z) \cdot {\tt v}({\tt v}_0,\lambda, \lambda_0) \cdot {{\rho \, {\rm cos} \theta}\over{\lambda}} \,\, dz} \over
{\int  W(\rho,\theta,z) dz}}
\label{rotcurve}
\end{equation}

\noindent
where $\lambda$ depends on $\rho$, $\theta$ and $z$ through Equation A1, and
the integral is performed over the range $-\sqrt{R_{\rm v}^2-\rho^2} < z < \sqrt{R_{\rm v}^2-\rho^2}$,
with $R_{\rm v}$ being the virial radius. For example, if we assume the velocity curve described in
Section 2, an isothermal ICM with a uniform metallicity, and a single $\beta$ model
for the ICM 3D density distribution, with $\beta=2/3$ for simplicity, we obtain the
following expression:

\begin{equation}
{{\tt v}_{\rm los}(\rho,\theta) = {\tt v}_0 c_{\tt v}\rho{\rm cos}\theta
\frac{\bigint_{-\sqrt{1-\rho^2}}^{\sqrt{1-\rho^2}}
\frac{dz}{(1+c^2(\rho^2+z^2))^2   (1+c_{\rm v} \sqrt{\rho^2{\rm cos}^2\theta+z^2})^2}   }
{\bigint_{-\sqrt{1-\rho^2}}^{\sqrt{1-\rho^2}} \frac{dz}{(1+c^2(\rho^2+z^2))^2} }}
\end{equation}

\noindent
where the two parameters describing the scale length are defined as $c=R_{\rm v}/R_c$ and
$c_{\tt v}=R_{\rm v}/\lambda_0$, and the variables $\rho$ and $z$ have been rescaled by the virial radius $R_{\rm v}$,
and therefore range in the interval $(0,1)$.
This formula is valid for a rotation axis on the plane of the sky and it has been used to
generate the velocity map in the left panel of Figure \ref{example}.  Maps with different rotation curves
can be obtained simply by substituting the curve function ${\tt v}({\tt v}_0,\lambda, \lambda_0)$.

\end{document}